# First-principles study of LiNb$_x$M$_{1-x}$O$_3$ (M= V, W, Ta, Mo,…) for holographic memory applications

S. Mondal[1], K. Choudhary[2], S. Pal[3], A. K. Bandyopadhyay[1], S. Das[4], and T. Chakraborty[5]

1) GCECT, W. B. University of Technology, Kolkata-700010, India

2) Department of Material Science & Engg, University of Florida, Gainesville, FL 32611, USA.

3) Metallurgical & Materials Engineering Department, National Institute of Technology, Rourkela-769008, India.

4) Department of Ceramic Engineering, IIT (BHU), Varanasi, Uttar Pradesh, 22005 India.

5) Department of Chemistry, Manipal University, Jaipur-303007, India

Corresponding Author: **asisbanerjee1000@gmail.com**

**Abstract:** For holographic memory applications, the photorefraction of well-known ferroelectric such as lithium niobate doped with different transition metals is very important. First principles study assumes special significance in this context, as to why certain transition metal atoms are better than the other. In this work, Nb atom in LiNbO$_3$ was substituted with transition elements having valency greater than/equal to +5 for UV photorefraction applications, and atomistic first-principles calculations were done using HSE06 functionals. The d-states of the transitional elements were found to decrease the band-gap of the host material having implications for a suitable material design. Minimum band-gap was obtained for W, while Ta showed a maximum value. Absorption coefficients were estimated for each material and based on their low values at 351 nm (i.e. for holographic applications) that is the usual UV photorefraction wavelength, the elements found suitable were V, W, Ta, Mo. Then birefringence properties for these crystals were also studied to predict that V and W were good candidates.

# I. Introduction:

Transition metal doped lithium niobate ferroelectric material is used for photorefraction and volume data-storage. Although a multitude of such materials have been characterized in details, the explanation as to why only a few of them are preferred compared to others is still to be known. In the field of condensed matter physics, one of the well-studied materials is ferroelectric, which has a variety of important applications in nonlinear optics, such as electro-optics and second harmonic generation, and non-volatile memory devices [1] (see references therein). Ferroelectrics have also emerged as important materials as piezoelectric transducers, pyroelectric detectors, surface acoustic wave (SAW) devices, four-phase mixing doublers. Such applications and many more as nano devices are mentioned in Ref [2-5]. One of the most important ferroelectric materials is lithium niobate (LN) that is widely used as an optical material for its various high frequency photonic and optoelectronics applications.

LN has gone through intensive research for its volume holographic data storage capability [6]. The theory of the volume phase holograms in the birefringent electro-optic LN crystals has been developed a long time ago via electrically controlled light diffraction and pattern reconstruction at a read wavelength different from that of the writing beams [7]. Holographic data storage is a technique that can store information at high density inside the crystals with the volume of the recording media used for data storage instead of just the surface. Opening up the 'inside' of a material for storage, rather than just using its surface, yields huge improvements in capacity. In addition, unlike conventional technologies, which record and recover digital data bit by bit, holography has the ability to read and write millions of bits of data with a single pulse of light, enabling data-transfer rates of billions of bits per second with a large storing capacity of several Tbs.

This subject remained dormant for quite some time. After a long break, the subject has been revived by an intensive activity on LN crystals doped with different transition elements **[8]**. The large crystal size of LN allows: (a) high capacity, (b) angular selectivity, (c) long storage lifetime, and (d) high diffraction efficiency. LN can be used in optical correlator type holographic data storage applications, where analog images are accessed at high speed. In particular, angle-multiplexing type storage in LN has been observed where 'optical correlation' is used for data storage **[9]**. Commonly Fe-doped LN is used for writing holograms in blue-green region of the visible spectrum, but in other parts of the electromagnetic spectrum, the 'sensitivity' is needed for inexpensive applications. In addition, LN:Fe has certain problems, such as (a) low response speed, (b) strong light-induced scattering, and (c) volatility. Other transition elements such as Cu, Mn, Ni, etc. have also been investigated for holography **[8]**. Early attempt of doping LN with more than one transition elements was made for the systems: LN:Fe,Mn and LN:Cu,Ce **[10,11]**. Also, there is an entire gamut of other doubly doped holographic materials based on LN, such as LN:Hf,Fe, LN:Zr,Fe, LN:Zr,Fe,Mn, LN:Hf,Fe,Mn, LN:Mn,Cu,Ce, Mn:Fe, Mn:Fe:Mg, Ln:Zr,Cu,Ce, which have also been tested for nonvolatile holographic storage. Mg, Zn, In, and Sc were found to be optical damage-resistant additives, and LN:Ru was found to have appreciable nonvolatile recording capacity **[12,13]** (see references therein for other important references on the above systems). Further improvement was made with LN:Hf, LN:Zr, LN:Sn, LN:Ru,Fe and LN:Zr:Ru:Fe **[14]**, and they were found to have similar optical damage-resistant effect compared to LN:Mg. Hence, among transition metals, the elements that were found to be suitable for holographic memory applications are: Sc, Zr, Hf, V,

Mo, Mn, Fe, Ru and Cu. It was also found that the substituents occupying Li-sites were those having valence less than 5+ **[12]**. Hence, others were found to occupy Nb sites and were primarily used for studying UV-photorefraction. Next, let us focus on the subject matter of this paper, i. e. density functional approach to tailor the use of different dopants.

Despite intensive experimental work, comparatively less theoretical investigation has been carried out on this issue, as well as on density functional theory (DFT) approach for atomistic descriptions. One of the pioneering investigations on the electronic structures and energetics of LN in comparison with those of another ferroelectric (LT) was done by Inbar and Cohen **[15]** via 'linearized augmented plane-wave' frozen phonon calculations in order to shed light on phase transition and origin of ferroelectricity. In this connection, an extensive amount of work done by various workers, such as Vanderbilt et al and Ghosez et al on first principle studies of different ferroelectrics need to be mentioned **[16]**. Now, for the electronic and optical properties of LN, various levels of DFT calculations such as general gradient approximation (GGA) and GW calculations were carried out on LN **[17]**. The band gap of LN was a crucial issue and was a much discussed topic of research in the last few decades [**18**]. Photorefractive property of both congruent and stoichiometric LN was investigated by many workers **[12,19]** that was related to birefringence property. Alkali metal doped LN was also studied with GGA theory by Aliabad and Ahmad **[20]**.

Considering the above, it is clearly evident that there is a necessity of studying the atomistic electronic properties, particularly in relation to band-gap engineering of such photorefractive materials involving LN via density functional theory (DFT) approach. This assumes more importance due to the recent surge in the activity in the field of holographic memory applications. In the present work, for atomistic calculations we study LN based materials involving different dopants that have been used (or possibly used) for photorefraction having valence more than/equal to 5+, namely V, Ta, Cr, Mo, W, Tc, Re, Ru, Os, Rh, Ir, Pt by using HSE06 hybrid functional. This is basically a DFT work related to holographic material exploration. In Section 2, we describe the method of calculations. In Section 3, we describe the results and discussion in terms of band-gap reduction, density of states as well as optical properties of LN with different substituents. Section 4 contains the conclusions.

## II. **Computational Details:**

For deciding on the doped LN, the system studied here is represented as $LiNb_xM_{1-x}O_3$, where M = V, Ta, (valency=+5), Cr, Mo, W, Tc, Ru, Os, Rh, Ir, Pt (valency=+6) and Re (valency=+7) respectively. The doping level considered here is at a quite high level at about 17 mol% for all the elements. Here, all the calculations have been carried out using density functional theory (DFT) involving Vienna Ab-inito Simulation Package (VASP) **[21]** within the first generalized gradient approximation (GGA), as formulated by Perdew, Burke and Ernzerhof **[22]**, and then hybrid functional (HSE) calculations were made. The projector-augmented wavefunction (PAW) approach, developed by Blochl, and adapted and implemented in VASP, was used for the description of the electronic wavefunctions. Plane waves have been included up to an energy cutoff 400 eV. Monkhorst-Pack 4x4x4 k-point mesh was used for the integration over the Brillouin zone. The energy criterion convergence was $10^{-6}$ eV.

$LiNbO_3$ with 20 atoms in a supercell with *a* = 5.15Å and *c* to be 13.865Å was relaxed with 4x4x4 k-points using GGA. Now, GGA is known to underestimate the band-gap of materials.

Hence, hybrid functional HSE calculations were carried out at 4x4x2 k-points. After relaxation *a* was found to be 5.13Å and *c* is 13.834Å, and then the optical properties were evaluated.

## III. Results and Discussion:

For holographic application, the lithium niobate system with different substituent atoms with valency 5+ and above was studied via DFT calculations. After presenting our results on band-gap reduction of LN via doping with different elements, as mentioned above, we direct our attention to an important aspect of microscopic properties, i.e. the density of states in LN with different substituents and the absorption spectra of these materials. At the end, we present some birefringence data.

### 3.1. Density of States (DOS):

Electronic distribution with respect to energy spectrum is shown by DOS that is experimentally determined by photoemission experiments. The DOS for pure LN is shown in Fig. 1a. The Fermi energy in all the plots is scaled to zero for convenience of interpretation and comparison. Band-gap can be determined from the DOS plots. For LN, the band gap was found to be 5.07 eV, which is justifiable with respect to experimental and previous theoretical studies. The valence bands are made mainly by O (2p) and slight covalent admixture of Nb (4d) states, while the conduction band is made by Nb (4d) with slight contribution due to O (2p) states. The valence band width is 5.4 eV, while the conduction band width is 2.2 eV. Nb (4p) states lie in the -32 eV region of the spectrum. There are no states found between 7 to 9 eV, after which the electronic states again populate. These states are mainly attributed to O (2s,2p), Nb (4d) and Li (2s) states.

After substituting Nb with the transition elements, the band-gap of the host-crystal was calculated. For the above dopant elements, in order to get an idea about the magnitude of reduction of indirect band-gap, it was calculated in eV (given in the bracket after each element), and was found to be quite large for Ta (5.05) that is almost similar to that of LN (5.07 eV). This is expected, as the structure of LN and LT is quite similar. For a set of elements with lower reduction in the band-gap (i.e. moderately higher band-gap), the results showed: V (3.77), Ir (3.66), Ru (3.48), Pt (3.37), Os (2.25) and Cr (2.04) respectively, while lower band-gap (i.e. higher reduction) was observed for Tc (1.66), Mo (1.57), Rh (1.36), Re (1.30) and remarkably the lowest band-gap was observed for W (0.72) indicating a possible application of the reduction of the 'band-gap' in some device material formulations. In the realm of band-gap engineering, the results clearly show that tantalum (Ta) has the minimum impact, while tungsten (W) has the maximum effect on the host crystal on the reduction of band-gap within the systems studied under the present investigation. Next, it is interesting to study the density of states (DOS) of such materials.

The total DOS is shown for the substituents in Fig. 1. The above mentioned decrease in band-gap is readily observed. The d-states of the substituent atoms are highlighted here, as they contribute mainly for this decrease. The decrease in the gap due to these states could also be viewed in relation to the nature of splitting of the conduction band minimum (CBM) and the valence band maximum (VBM). The conduction band (CB) is split for V, Cr and Pt, while the valence band (VB) is split for Mo, Tc, Ru, Rh, W, Re, Os, Ir and Pt. In Fig. 1, both the spin-up and spin-down data are shown. Significantly, it is seen that the spin-up and spin-down components are not aligned for Cr, Mo, Tc, Rh, W and Re respectively that might lead to spin-up and spin-down transitions having implications in the relevant magnetic properties of such materials. It is pertinent to mention about an interesting work by Tong et al. **[23]** (see references therein) who observed significantly different optical responses from spin-up and spin-down channels in case of multiferroic EuO via first principles DFT calculations. To remain within our main focus area, let us look at the absorption coefficients.

### 3.2. <u>Absorption Spectra:</u>

It is an important tool or technique in designating optical properties of materials. The data are shown in Fig. 2. It is obtained from the dielectric function based on inter-band transitions and symmetry constraints. For pure LN, the absorption spectrum could be interpreted with the help of DOS. The peaks in the 5.07 to 12.5 eV are due to valence to conduction states. The higher energy states are due to the bands that are composed by Li (s) and O (p) states. After introducing the substituents, the states are added in the gap region as well as at the bottom of valence band and the top of the conduction band, introducing absorption peaks within 5.07 eV and modifying the UV-characteristics. The characteristics in the *x* (dashed line) and *z* (solid line) directions are not the same that obviously gives an indication about birefringence. The multiple DOS peaks in the gap region lead to various peaks in the absorption spectrum that is observed in this gap region for various elements, such as for Rh, Ru, Tc, Os, Ir and Pt respectively, except that Ta-substituted LN shifts the band made by Nb (p) and O (s) states deeper in the energy spectrum, leading to shifting of absorption peaks around 20 eV.

For holographic experiments, a 351 nm laser is generally used for measuring the photorefraction. Therefore, the absorption coefficients were monitored at 351 nm (~3.53 eV). Low values (in $10^{-4}$ cm$^{-1}$) were observed at 351 nm for some elements. In the order of decreasing absorption coefficient, the results for eight elements showed: Rh (1.1), Ir (0.9), Re (0.8), Pt (0.7),

Mo (0.43), V (0.42), W (0.2), and Ta (0.1) respectively, thus making Ta, W, V, Mo the possible candidates as Nb substituents in LN. Starting from the lowest value with LN:W at 0.72 eV, a comparison of the band-gap of these doped materials vis-à-vis their absorption coefficients shows that the absorption first peaks at 1.36 eV for LN:Rh and then decreases to show up another maximum again at much higher band-gap of 3.66 eV for LN:Ir. In a non-exhaustive literature survey, in the absence of band-gap data for such a wide variety of substituents in LN for holographic applications, it is quite difficult to interpret these results. However, among the four chosen elements as mentioned above, Ta is very well-known as nonlinear optical material showing the lowest absorption coefficient for LN:Ta at 0.1, although its band-gap is the highest (5.05 eV). However, for LN:W whose absorption coefficient is also very low at 0.2, its band-gap is the lowest (0.72 eV).

This opens up a new field of work in the realm of band-gap engineering of device material formulations to suit different applications. For example, in a recent work by Grinberg et al **[24]**, a 10% replacement of $KNbO_3$ by nickel-niobium based material decreased the band-gap from about 4.2 eV to 1.39 eV to give rise to a very high level of photovoltaic efficiency in terms of high photocurrent density for solar-energy conversion and other important applications for this type of ferroelectric-semiconductor based materials. It is pertinent to mention that in the present investigation with our DFT approach, leaving aside LN:W, a comparable band-gap was observed for LN:Re and LN:Rh and LN;Mo at 1.30 eV, 1.36 eV and 1.57 eV respectively. These materials could also be considered as competitive candidates for similar applications. In this context, the birefringence of such doped materials assumes some significance that was also calculated from the present DFT approach. Hence, we next present the birefringence of these materials in Fig. 3. It is seen that the lowest value was observed for V and the highest for Ta doping. Birefringence is directly related to the photorefraction properties. It varies with the incident beam energy. Below 3.5 eV, W has the lowest birefringence, but after this value, V is lower than W. Then at higher energy, upto 4.3 eV, Mo has the lowest birefringence. Further, both W and Ta attain low values at high energy. It seems to give a certain direction on the right type of material that could possibly be used for a particular energy beam.

No photorefraction measutrments have been performed on the above cited doped LN materials. However, it is clearly seen that doping helps in reducing the band-gap to a significant extent, as revealed by atomistic simulation work via DFT approach, and it opens a new field of applications for both memory (holographic) as well as for solar-energy generation.

## IV. Conclusions:

Nb atom in $LiNbO_3$ was substituted with transition elements having valency greater than or equal to +5 for UV photorefraction applications in the domain of holographic memory systems. A detailed calculation was made with DFT approach using HSE06 functionals. The d-states of the transitional elements were found to decrease the band-gap of the host material. Minimum band-gap, but a very significant reduction was obtained for W at 0.72 eV, while maximum value was attained in case of Ta at 5.05 eV. The absorption coefficients were evaluated for each material and based on its low values at 351 nm, which is the 'working wavelength of UV photorefraction, the elements found to be suitable were W, Mo, V and Ta respectively in increasing order of band-gap. In the band-gap of these doped entities, the substituent elements, i.e. Ta and W, with extreme values (5.07 eVand 0.72 eV) show the lowest absorption coefficients, i.e. 0.1 and 0.2 respectively. Then, the birefringence properties for these crystals were also evaluated. The lowest birefringence was obtained for V.

**Acknowledgements:** The authors would like to thank Arthur McGurn of Physics Dept of Western Michigan University (USA) for helpful comments.

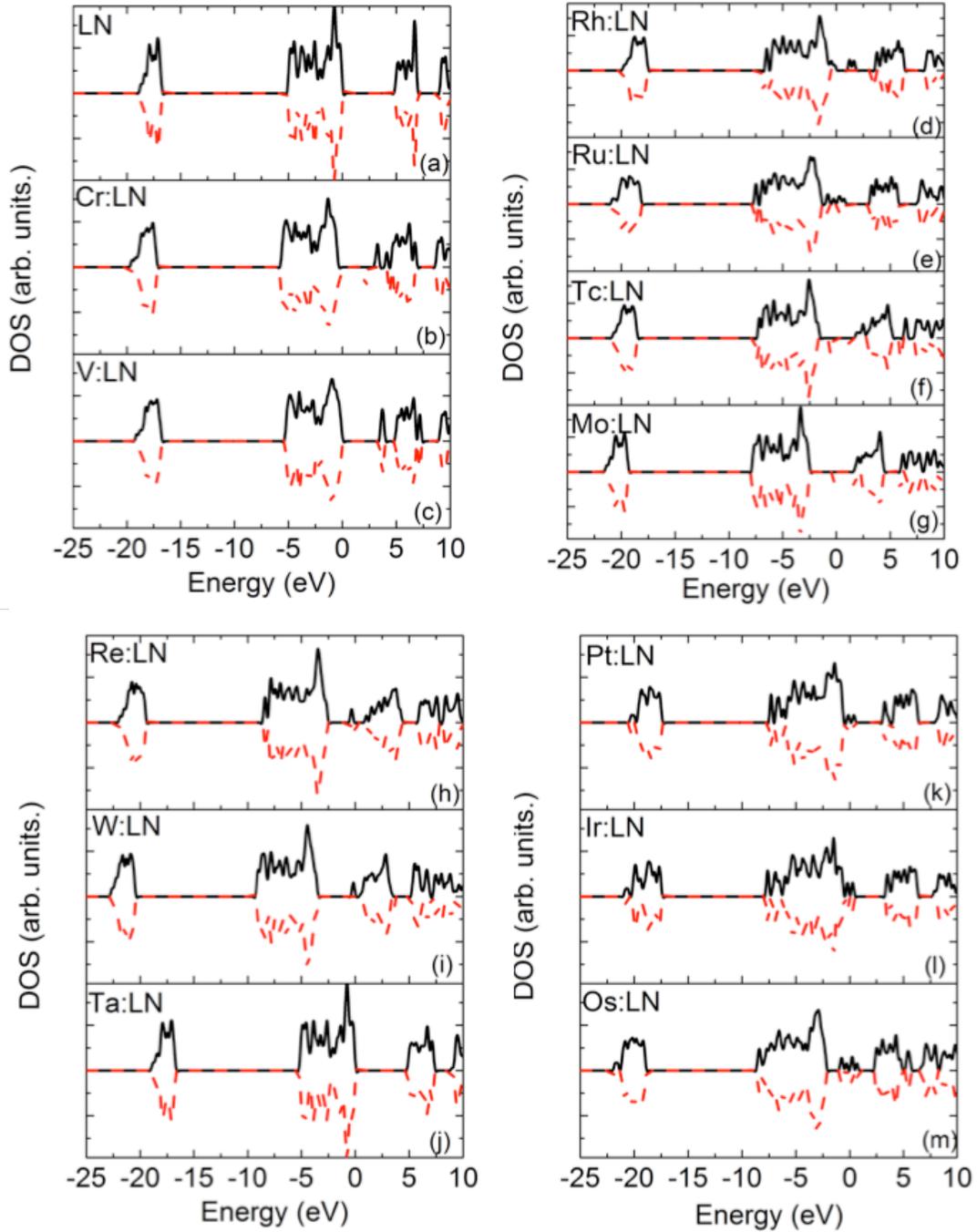

Fig. 1: Components of Spin-up (upper portion) and Spin-down (lower portion) of total DOS.

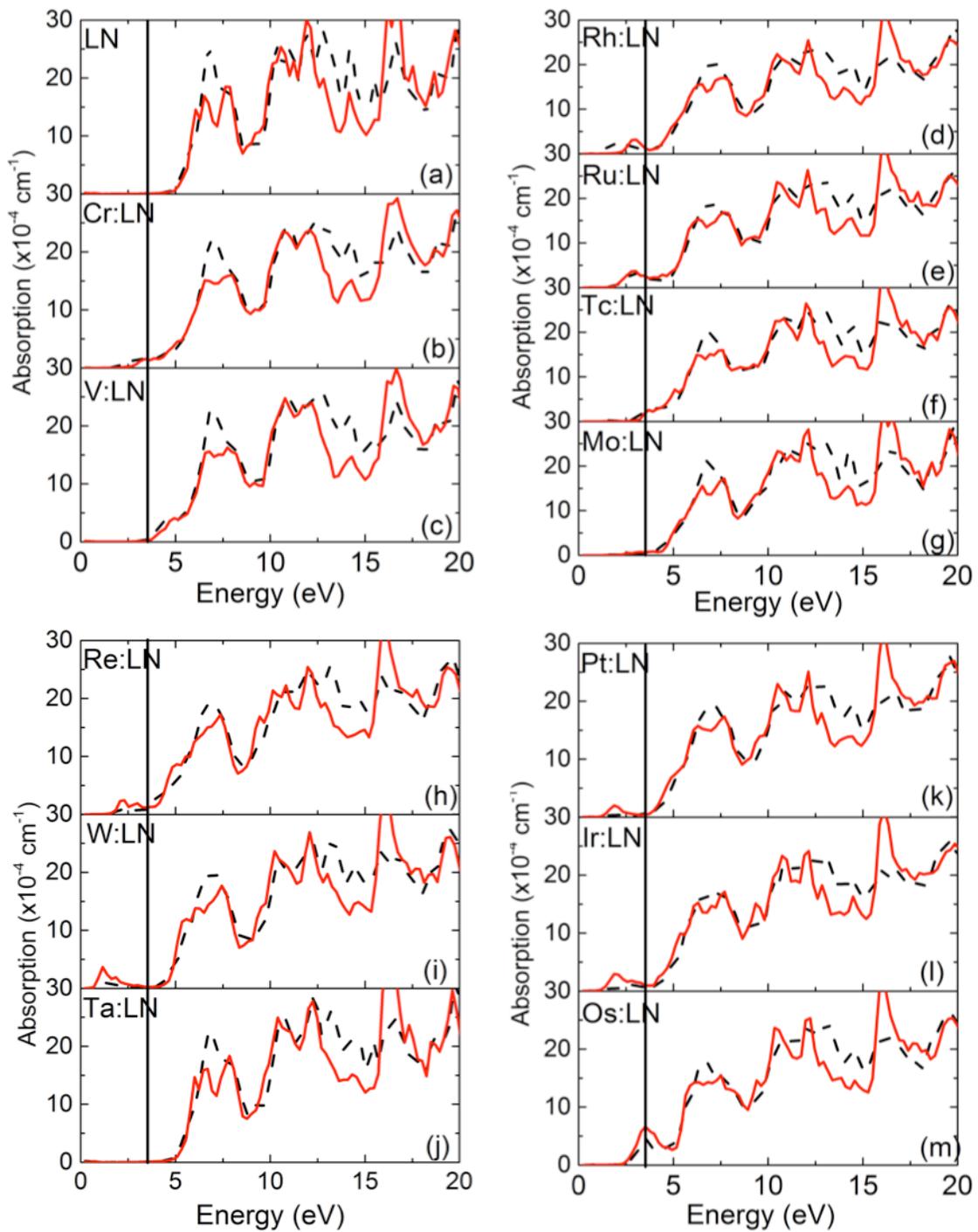

Fig. 2: Absorption spectra. A reference line is given at 3.53 eV, which is the usual wavelength for holographic experiment.

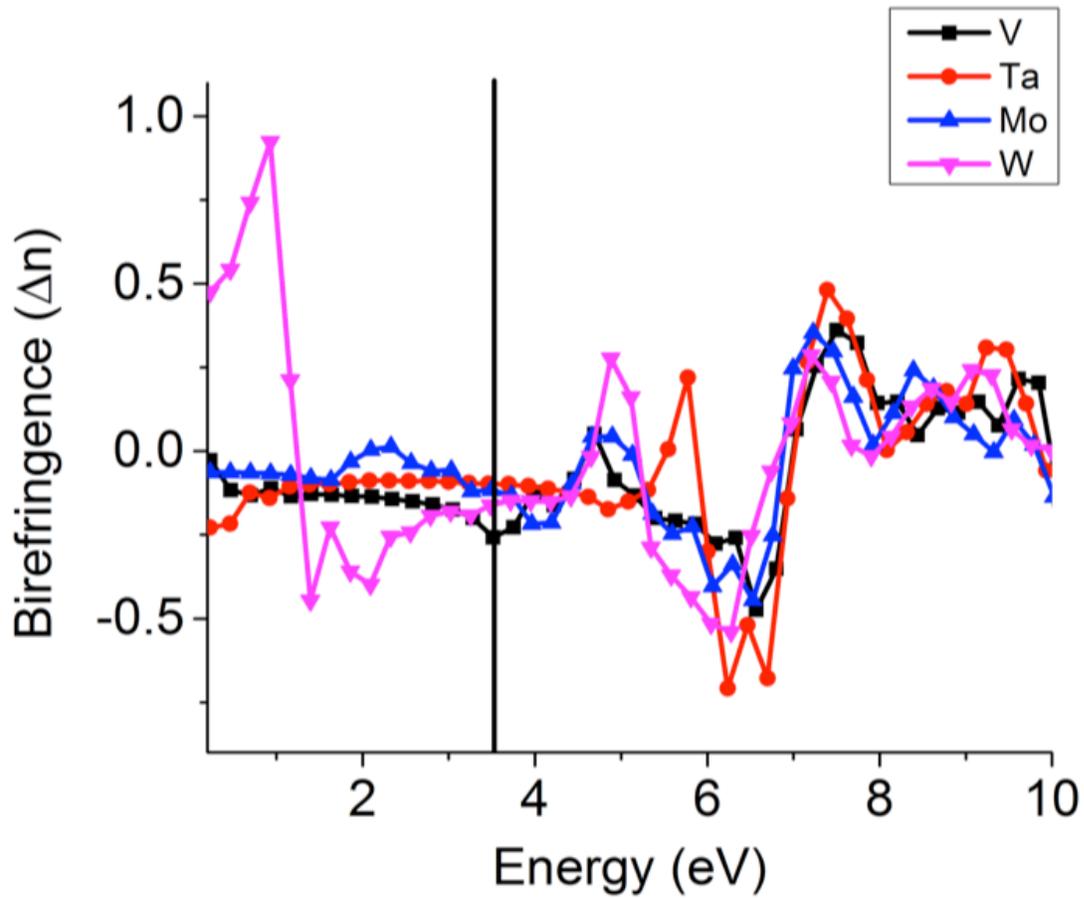

Fig. 3: Birefringence of V, W, Ta, Mo doped LN. A reference line is given at 3.53 eV.